\documentstyle[prl,aps]{revtex}
\def\be{\begin{equation}}
\def\ee{\end{equation}}
\def\bea{\begin{eqnarray}}
\def\eea{\end{eqnarray}}
\def\ba{\begin{array}}
\def\ea{\end{array}}

\begin{document}
\title{Octet magnetic moments and the violation of CGSR
in $\chi$QM with configuration mixing}
\author{Harleen Dahiya  and Manmohan Gupta}
\address {Department of Physics,
Centre of Advanced Study in Physics,
Panjab University,Chandigarh-160 014, India.} 
 \maketitle

\begin{abstract}
Octet baryon magnetic moments are calculated within $\chi$QM, respecting
color spin spin forces (Szczepaniak {\it et al.}, PRL
{\bf 87}, 072001(2001)), incorporating the
orbital angular momentum as well as the quark sea contribution 
through the Cheng and Li mechanism (PRL {\bf 80}, 2789(1998)). Using
configuration mixing generated by color spin spin forces as well as the
concept of ``effective'' quark mass to include the effects of confinement, 
we are able to get an excellent fit to the octet magnetic moments as
well as the violation of
Coleman Glashow Sum Rule (CGSR) without any further input except for
the ones already used in $\chi$QM as well as in NRQM.
Specifically, in the  case of $p, \Sigma^+$, $\Xi^o$, and violation of
CGSR we 
get a perfect fit whereas in almost all the other cases the results 
are within 5\% of the data. 
\end{abstract}
 
%\pacs{}

The EMC  measurements \cite{EMC} in the deep inelastic scattering
had shown that only a small fraction of the proton's spin is carried by the
valence quarks. This ``unexpected'' conclusion from the point of view of 
Non Relativistic Quark Model (NRQM), usually referred to as 
``proton spin crisis'',
becomes all the more intriguing
when it is realized that NRQM is able to give a reasonably
good description of magnetic moments using the assumption
that magnetic moments of quarks are proportional to the spin carried
by them. Further, this issue regarding spin and magnetic moments becomes 
all the more difficult to
understand when it is realized that the magnetic moments of
baryons receive contribution not only from the magnetic moments carried
by the valence quarks but also from various complicated effects,
such as, orbital excitations \cite{orbitex}, 
sea quark polarization \cite{{eichten},{song},{cheng}},
effects of the chromodynamic spin-spin forces 
\cite{{Isgur},{mgupta1}}, effect of the confinement
on quark masses \cite{{effm1},{effm3}}, 
pion cloud contributions \cite{{pioncloudy}}, 
loop corrections \cite{loop},
relativistic and exchange current effects \cite{excurr}, etc. 
The problem regarding 
magnetic moments gets further 
complicated when one realizes that
Coleman Glashow Sum Rule (CGSR) \cite{cg}, valid in large variety of models
\cite{{cg1},{johan}}, is convincingly violated by the 
data \cite{deltacg}. If $\Delta$CG is the deviation from the
CGSR, then experimentally
$\Delta{\rm CG}=0.49 \pm 0.05,$
clearly depicting the violation of CGSR by ten standard deviations.
As $\Delta$CG=0, in most of the calculations, a reproduction of
$\Delta$CG would provide a viable clue for the dynamics of
constituents of nucleon.

In this context, it is interesting to note that the 
Chiral Quark Model ($\chi$QM) {\cite{{cheng},{manohar}}}
with SU(3) symmetry is not only able to give a fair explanation of
``proton spin crisis'' {\cite{EMC}} but is also able to give 
a fair account of $\bar u-\bar d$ asymmetry {\cite{{NMC},{E866},{GSR}}} as well
as the existence of significant strange
quark content $\bar s$ in the nucleon {\cite{st q}}.
Further, $\chi$QM with SU(3) symmetry is also able to provide fairly 
satisfactory explanation for  baryon magnetic moments 
{\cite{{eichten},{cheng},{manohar}}} 
as well as the absence of polarizations of 
the antiquark sea  in the nucleon {\cite{antiquark}} .
The predictions of the $\chi$QM, particularly in regard to  
hyperon decay parameters\cite{decays}, 
can be improved if symmetry breaking effects \cite{song} are
taken into account. However, $\chi$QM with symmetry 
breaking, although gives a fairly good description of magnetic moments, 
is not able to describe $\Delta$CG.
In this regard, it is to be noted that some recent attempts to describe 
$\Delta$CG within $\chi$QM, resort to additional inputs 
\cite{{johan},{cgv1}} over and above the basic premises of 
$\chi$QM.

In a recent interesting  work, Cheng and Li \cite{cheng1} have shown 
that, within  $\chi$QM with SU(3) symmetry breaking, a  long standing puzzle, ``Why the NRQM
is able to give a fair description of baryon magnetic moments''
can be resolved if one considers the pions acting as Goldstone Bosons 
also have angular momentum and therefore, contribute
to the baryon magnetic moments as well. However, this contribution gets
effectively cancelled by the sea quark polarization effect leaving the
description of magnetic moments in terms of the valence quarks in
accordance with NRQM hypothesis. 
One can easily examine that the Cheng and Li proposal does not lead to exact
cancellation of the sea and orbital part for all the baryons,
therefore it needs to be examined in detail for the 
octet baryon magnetic moments.

Similarly, in a  very recent development,
Szczepaniak and  Swanson \cite{prl} have shown that the 
constituent quarks undergoing
hyperfine chromodynamic spin spin interactions respect chiral symmetry.
In view of the fact that $\chi$QM incorporates constituent quarks as
essential ingredients and NRQM with chromodynamic spin spin forces 
is known to describe vast amount of spectroscopic data 
\cite{{prl},{yaouanc},{DGG},{photo}}, it therefore becomes 
interesting to investigate the problem of magnetic moments in $\chi$QM with 
spin spin forces.
Infact, it has already been shown recently \cite{hd} that the 
$\chi$QM with spin spin forces ($\chi$QM$_{gcm}$) as well as SU(3) and
axial U(1) symmetry breakings improves the predictions of 
$\chi$QM with SU(3) symmetry breaking, 
however the issue of magnetic moment has not been studied in detail.

One of the difficulty which NRQM has to face is how to include the
effects of quark confinement. Pending solutions of QCD in the
confinement regime, it has been shown that the effects of confinement
can be incorporated in NRQM by considering ``effective'' quark masses
\cite{{effm1},{effm3}}. In the context of magnetic moments
this essentially replaces constituent quark masses by ``effective''
quark masses, specific to a given baryon,  and it has been shown to
improve the performance of NRQM as far as magnetic moments are
concerned \cite{{effm1}}.

The purpose of the present communication is to investigate, within
$\chi$QM, the effects of orbital currents, as advocated
by Cheng and Li \cite{cheng1}, sea polarization, 
chromodynamic spin spin forces,
``effective'' masses on octet magnetic moments and CGSR.
While pursuing this goal, we have kept in mind that the
parameters responsible for successes of the $\chi$QM as well as the
NRQM with chromodynamic spin spin forces are not to be disturbed.

To begin with we consider the essentials of
$\chi$QM with chromodynamic spin spin forces 
($\chi$QM$_{gcm}$) discussed in detail in Ref \cite{hd}.  
In $\chi$QM, the basic process is the
emission of a Goldstone Boson (GB) which further splits into $q \bar q$
pair, for example,                         

\be
  q_{\pm} \rightarrow GB^{0}
  + q^{'}_{\mp} \rightarrow  (q \bar q^{'})
  +q_{\mp}^{'}.                              \label{basic}
\ee
The effective Lagrangian describing interaction between quarks
and the octet GB and singlet $\eta^{'}$ is
${\cal L} = g_8 \bar q \phi q,$
where $g_8$ is the coupling constant
and
\[ \phi = \left( \ba{ccc} \frac{\pi^o}{\sqrt 2}
+\beta\frac{\eta}{\sqrt 6}+\zeta\frac{\eta^{'}}{\sqrt 3} & \pi^+
  & \alpha K^+   \\
\pi^- & -\frac{\pi^o}{\sqrt 2} +\beta \frac{\eta}{\sqrt 6}
+\zeta\frac{\eta^{'}}{\sqrt 3}  &  \alpha K^o  \\
 \alpha K^-  &  \alpha \bar{K}^o  &  -\beta \frac{2\eta}{\sqrt 6}
 +\zeta\frac{\eta^{'}}{\sqrt 3} \ea \right). \]

SU(3) symmetry breaking is introduced by considering
different quark masses $m_s > m_{u,d}$ as well as by considering
the masses of GBs to be non-degenerate
 $(M_{K,\eta} > M_{\pi})$ {\cite{{song},{johan}}}, whereas 
  the axial U(1) breaking is introduced by $M_{\eta^{'}} > M_{K,\eta}$
{\cite{{song},{cheng},{johan}}}.
The parameter $a(=|g_8|^2$) denotes the transition probability
of chiral fluctuation
of the splittings  $u(d) \rightarrow d(u) + \pi^{+(-)}$, whereas 
$\alpha^2 a$, $\beta^2 a$ and $\zeta^2 a$ 
denote the probability of transition of
$u(d) \rightarrow s  + K^{-(0)}$,
$u(d,s) \rightarrow u(d,s) + \eta$ and
$u(d,s) \rightarrow u(d,s) + \eta^{'}$ respectively.

The valence quarks of the $\chi$QM undergo chromodynamic spin spin
forces as a consequence of which the constituent quark wave functions
get ``perturbed''. For the present purpose it has been shown
 {\cite{{mgupta1},{yaouanc}} that it is adequate to consider
\begin{equation}
\left|8,{\frac{1}{2}}^+ \right> = {\rm cos} \phi |56,0^+>_{N=0}
+ {\rm sin} \phi|70,0^+>_{N=2},  \label{mixed}
\end{equation}
for details of the perturbed wave functions, we refer the readers to
{\cite{{mgupta1},{yaouanc}}.

Using the above wave function along with the orbital and sea contributions, as 
discussed by Cheng and Li, and the ``effective'' quark masses, we have 
carried out the calculations of octet baryon magnetic moments. 
Including the corrections induced by the spin spin forces and
``effective'' quark masses, the magnetic moment corresponding to a given
baryon(B) can be written as
\be
\mu_{total} = \mu_{val} + \mu_{sea} +\mu_{orbit}, \label{totalmag}
\ee
where 
$\mu_{val}=\sum_{q=u,d,s} {\Delta q_{val}\mu_q}$,
$\mu_{sea}=\sum_{q=u,d,s} {\Delta q_{sea}\mu_q}$,    
$\mu_q$ ($q=u,d,s$) is the quark magnetic moment  and $\Delta q$ 
($q=u,d,s$) represents the net spin
polarization  and is defined as
$\Delta q= q_{+}- q_{-}+
\bar q_{+}- \bar q_{-}.$

The valence contribution, $\Delta q_{val}$, can easily be calculated
\cite{{song},{cheng},{johan},{hd}}, for example, using Eq (\ref{mixed}), for
proton we have, $\Delta u_{val} ={{\rm cos}}^2 \phi \left[\frac{4}{3} \right]
   + {{\rm sin}}^2 \phi \left[\frac{2}{3}  \right],~~
  \Delta d_{val} ={{\rm cos}}^2 \phi \left[-\frac{1}{3} \right]  +
  {{\rm sin}}^2 \phi \left[\frac{1}{3}  \right], ~~
   \Delta s_{val} = 0$. Similarly one can calculate for other baryons.
 The sea contribution in the $\chi$QM basically comes from the splitting
 of GB into $q \bar q$ pair (Eq (\ref{basic})).
The contribution to baryon magnetic moments from the sea can easily be 
calculated within $\chi$QM$_{gcm}$ \cite{{cheng},{johan},{hd}}. 
For the case of proton it is given as
\bea
\Delta u_{sea}&=&-{{\rm cos}}^2 \phi \left[\frac{a}{3} (7+4 \alpha^2+
 \frac{4}{3}\beta^2 +\frac{8}{3} \zeta^2)\right]
-{{\rm sin}}^2 \phi \left[\frac{a}{3} (5+2 \alpha^2
+\frac{2}{3}\beta^2 +\frac{4}{3} \zeta^2)\right], \\
\Delta d_{sea}&=&-{{\rm cos}}^2 \phi \left[\frac{a}{3} (2-\alpha^2
-\frac{1}{3}\beta^2 -\frac{2}{3} \zeta^2)\right]
-{{\rm sin}}^2 \phi \left[\frac{a}{3} (4+\alpha^2
+\frac{1}{3}\beta^2 +\frac{2}{3} \zeta^2)\right] \\ 
 \Delta s_{sea}&=&-a \alpha^2.
\eea 
Similarly one can calculate $\Delta q_{sea}$ and 
consequent contribution to magnetic moments  for other baryons.

Following Cheng and Li  \cite{cheng1}, the $\mu_{orbit}$ for 
$\chi$QM$_{gcm}$ can easily be evaluated and for proton it is given as
\be
\mu_{orbit} =\left[ \frac{4}{3} {{\rm cos}}^2 \phi
+\frac{2}{3}{{\rm sin}}^2 \phi  \right] \left[\mu (u_+ \rightarrow) \right]+
\left[ -\frac{1}{3}{{\rm cos}}^2 \phi  +
\frac{1}{3}{{\rm sin}}^2 \phi \right] \left[\mu (d_+ \rightarrow)
\right],  \label{orbit}
\ee
 where $\mu(u_+ \rightarrow)$ and $\mu(d_+ \rightarrow)$ are the orbital
moments of $u$ and $d$ quarks and are given as
\bea
\mu(u_+ \rightarrow) &=& \frac{a}{2 {M}_{GB}(M_u+{M}_{GB})}
\left[ 3(\alpha^2+1)M_u^2 + (\frac{1}{3}\beta^2+\frac{2}{3}\zeta^2 -
\alpha^2){{M}_{GB}}^2 \right]{\mu}_N, \label{u} \\
\mu(d_+ \rightarrow) &=& \frac{a}{4 {M}_{GB}(M_u+{M}_{GB})} \frac{M_u}{M_d}
\left[ (3- 2 \alpha^2-\frac{1}{3}\beta^2-\frac{2}{3}\zeta^2)
{{M}_{GB}}^2 -6 M_d^2 \right]{\mu}_N,     \label{d} \\
\eea
($M,M_{GB}$) are the masses of quark and GB, 
$\mu_N$ is the Bohr magneton. In a similar manner one can calculate the 
contributions for other baryons.

Before  discussing  the results, first we discuss the inputs required
for numerical calculations. As is evident from the Eqs
(\ref{totalmag}),
the different inputs required pertain
to $\chi$QM, configuration mixing generated by the chromodynamic spin
spin forces and the NRQM. 
To this end, we have used the  $\chi$QM parameters  $a=0.1$, $\alpha=0.4$
and $\beta=0.7$, as used in the case of  $\chi$QM$_{gcm}$
\cite{hd}. However, to fit the  violation of the
Gottfried Sum Rule \cite{GSR} we have used
$\zeta=-0.3-\beta/2$ and $\zeta=-0.7-\beta/2$ respectively for  the
E866 and the NMC data.

The orbital angular moment contributions  are  characterized by
the parameters of $\chi$QM as well as the masses of the GBs. In view
of the fact that the orbital contributions are dominated by the pion  
contributions, the contributions of other GBs 
being much smaller as compared to pionic contributions  have
been ignored  in the numerical calculations. For evaluating the contribution 
of pions we have used its on mass shell value 
in accordance with several other similar calculations 
\cite{{mpi1}}. 

In the absence of any definite guidelines for the constituent quark
masses, we have used  their most widely accepted values in hadron
spectroscopy 
\cite{{yaouanc},{DGG},{photo},{close}}, 
for example, $M_u=M_d=330$ MeV.
The strange quark mass corresponding to a given baryon 
is fixed from the
sum rules implied by the chromodynamic spin spin forces, for example,
$ \Lambda-N =M_s-M_u$, $(\Sigma^*-\Sigma)/(\Delta-N)=M_u/M_s$ and 
 $(\Xi^*-\Xi)/(\Delta-N)=M_u/M_s$ respectively fix 
$M_s$ for $\Lambda$, $\Sigma$ and $\Xi$ baryons. 
The above mentioned quark masses and corresponding magnetic moments
have to be further corrected by the confinement effects 
\cite{{effm1}} which are effectively simulated in the present
context by considering the quark magnetic moments as 
$\mu_d = -\left(1-\frac{\Delta M}{M_B}\right) {\mu}_N,~
\mu_s = -\frac{M_u}{M_s} \left(1-\frac{\Delta M}{M_B}\right) {\mu}_N,
~\mu_u=-2 \mu_d$, where
$M_B$ is the mass of the baryon obtained
additively from the quark masses and $\Delta M$
is the mass difference between the experimental value and $M_B$.

In Table 1, we have presented the results of our calculations for the
octet baryons respectively for E866 and NMC data.
From  the table, it is
evident that with the case of E866 data 
we are able to get an excellent fit for almost all the
baryons. 
In fact, the fit is almost perfect for $p, \Sigma^+$ and $\Xi^o$,
in the case of $n, \Sigma^-$ and $\Lambda$ 
the value is reproduced within 5\% of
experimental data. Only in the case of $\Xi^-$ the deviation is
somewhat more than 5\%. 
Besides this we have also been able to get an excellent fit to the
$\Delta$CG. 
The fit becomes all the more impressive when it is
realized that none of the magnetic moments are used as inputs. 
The results are equally impressive in the case of NMC data, however
the lower values of individual magnetic moments compared to E866 data
can be attributed to lower value of 
$|\zeta_{E866}|$ as compared to $|\zeta_{NMC}|$.

In the table, we have also mentioned our results without the
configuration mixing, primarily to understand
the role of configuration mixing. From the table we find that
without configuration mixing the results are generally on the higher side
which get corrected in the right direction by the inclusion of
configuration mixing, for example, the valence contribution
of the proton reduces form 3.17 to 2.94 and that of $\Sigma^+$
from 2.80 to 2.59.
It is  noted that configuration mixing reduces
valence, sea and orbital contributions to the magnetic moments,
however in the case of $\Xi^-$ the configuration mixing improves the
fit by contributing with the right sign.
Thus, it seems that for the quark masses, which are able to
reproduce hadronic spectroscopy, configuration mixing is
important to fit individual magnetic moments.

A closer scrutiny of our results reveals several interesting
points. The most interesting point is that the orbital angular
momentum contribution
and the contribution by sea quarks are fairly substantial, however they
appear with the opposite sign in accordance with the conclusion of 
Cheng and Li \cite{cheng1}. In fact these contributions are delicately
balanced so as to improve the overall fit of the magnetic moments
without disturbing the successes of $\chi$QM. 
In fact from the table one can easily find out that except
for $\Xi^-$, in all other cases the valence+sea contribution
contributes to the overall magnetic moments with the right sign. This
strongly supports the Cheng and Li dynamics for the constituents of
the nucleon. One may wonder how to fit the magnetic moment of $\Xi^-$
in the present scheme. As mentioned earlier, there are several
other effects such as pion cloud contributions \cite{pioncloudy}, 
loop corrections \cite{loop},
relativistic and exchange current effects \cite{excurr} etc. 
which also affect
the magnetic moments. We believe the inclusion of these effects would
perhaps improve the present fit further. In fact a cursory look at
\cite{pioncloudy} suggests that pion loop corrections would compensate
$\Xi^-$ much more compared to other baryons hence providing an almost
perfect fit.

It should also be mentioned that the mixing generated by spin spin
forces and mass correction induced by confinement effects also play an
important role in achieving an overall fit. In fact, it is the mass
correction which affects the valence contribution to the magnetic
moments with the right magnitudes in getting the right value of
$\Delta$CG. 

To summarize, baryon magnetic moments have been calculated in the
$\chi$QM, with one gluon exchange generated configuration mixing,
 by including the contribution of the
$q \bar q$ sea polarization, orbital contribution of the sea as well
as the correction induced on quark magnetic moments due to confinement
effects. The calculations have been carried out by employing the same
parameters which are responsible for the successes of $\chi$QM, in
explaining the quark and spin distribution functions, as well as those
of NRQM responsible for explaining vast amount of spectroscopic
data. Apart from getting an excellent
fit to the baryon magnetic moments, we are also able to get an almost
perfect fit for $\Delta$CG.
In fact, the fit is almost perfect for $p, \Sigma^+$ and $\Xi^o$,
in the case of $n, \Sigma^-$ and $\Lambda$ 
the value is reproduced within 5\% of
experimental data. Only in the case of $\Xi^-$ the deviation is
somewhat more than 5\%. 
On closer examination of the results we find configuration mixing,
effect of confinement on quarks as well as the orbital and sea
polarizations, all play a crucial role in effecting the fit. In fact
the mismatch in the orbital and sea polarizations correct the magnetic
moments of valence quarks in the right direction in almost all the
cases. This strongly supports the Cheng and Li dynamics for the $\chi$QM.
 
 The authors would like to thank S.D. Sharma for fruitful 
discussions.
H.D. would like to thank CSIR, Govt. of India, for
 financial support and the chairman,
 Department of Physics, for providing facilities to work 
 in the department.

\begin{table}
{\footnotesize
\begin{center}
\begin{tabular}{cccccccccccccccc}      
 &  & \multicolumn{7}{c} {$\chi$QM without configuration mixing} &
\multicolumn{7}{c} {$\chi$QM with configuration mixing}\\ \cline{3-16} 

Octet & Data &   Valence  & \multicolumn{2}{c} {Sea} & 
\multicolumn{2}{c} {Orbital}
&  \multicolumn{2}{c} {Total} &  Valence & \multicolumn{2}{c} {Sea}
& \multicolumn{2}{c} {Orbital} &  \multicolumn{2}{c} {Total} \\   
 baryons & \cite{deltacg} &     & \multicolumn{2}{c} {} & 
\multicolumn{2}{c} {}
&  \multicolumn{2}{c} {} &  & \multicolumn{2}{c} {}
& \multicolumn{2}{c} {} &  \multicolumn{2}{c} {} \\   
&&& E866 &NMC & E866 &NMC & E866 &NMC && E866 &NMC & E866 &NMC & E866 &NMC \\
\hline

p          & 2.79  &   3.17   &  -0.59&-0.73  &  0.45 &0.44&  3.03 &2.88&
       2.94    & -0.55&-0.69   & 0.41 &0.40  & 2.80& 2.65   \\
n          & -1.91 &  -2.11    & 0.24  &0.32 & -0.37 & 0.38& -2.24 & -2.17&
 -1.86  &  0.20 &0.27& -0.33 &-0.34 & -1.99&-1.93    \\

$\Sigma^-$ & -1.16 &   -1.08 & 0.08&0.12  & -0.26 &-0.27&   -1.26 &-1.23&
   -1.05    & 0.07 &0.14 &  -0.22 &-0.26 &  -1.20&-1.17   \\
$\Sigma^+$ & 2.45  &  2.80  &  -0.55 &-0.68&   0.37 &0.39& 2.62 &2.51& 
 2.59 & -0.50 & -0.64& 0.34 &0.36&  2.43 &2.31 \\

$\Xi^o$    & -1.25 &  -1.53 & 0.22 & 0.30& -0.16 &-0.17 &  -1.47 &-1.40& 
 -1.32 & 0.21 &0.26& -0.13 &-0.14&  -1.24 &-1.20   \\
$\Xi^-$    & -0.65 &  -0.59 & 0.06  &0.09 &  -0.01 &-0.01 &  -0.54 &-0.51&  
 -0.61 & 0.06& 0.09  & -0.01 & -0.02 &  -0.56&-0.54  \\

$\Lambda$ & -0.61  & -0.69  & 0.05 &0.10   & -0.04& -0.05  &-0.68  & -0.64&
-0.59 & 0.04&0.07  &-0.04& -0.05  & -0.59& -0.57   \\
 \hline
$\Delta$CG  & 0.49 $\pm$ 0.05 &   &&&  &  & 0.46 & 0.42 & 
 &&&  &  &  0.48& 0.44  \\

\end{tabular}
\end{center}}
\caption{ Octet baryon magnetic moments (in units of $\mu_N$).} 
\end{table}
\end{document}